\documentclass[12pt,letterpaper]{article}

\usepackage[margin=1in]{geometry}
\usepackage{amsfonts}
\usepackage{amsmath}
\usepackage{amssymb}
\usepackage{graphicx}
\usepackage{times}
\usepackage{paralist}
\usepackage{bbm}
\usepackage[FIGTOPCAP]{subfigure}
\usepackage[font=small]{caption}
\usepackage{booktabs}
\usepackage{comment}
\usepackage{bm}
\usepackage{color}
\usepackage{setspace}
\usepackage{lineno}
\usepackage{url}

\newcommand{\R}{\ensuremath{\mathbb{R}}}
\newcommand{\bfx}{\ensuremath{\mathbf{x}}}

\newcommand{\E}{\ensuremath{\mathbb{E}}}
\newcommand{\Cov}{\ensuremath{\mathrm{Cov}}}
\newcommand{\Var}{\ensuremath{\mathrm{Var}}}

\newcommand{\bl}{\begin{linenomath}} 
\newcommand{\el}{\end{linenomath}}

\usepackage{natbib}
\title{\bf Assessing the Calibration of High-Dimensional Ensemble Forecasts Using Rank Histograms}

\author{Thordis L. Thorarinsdottir\footnote{Norwegian Computing Center, Oslo, Norway. {\em Corresponding author: thordis@nr.no}}, Michael Scheuerer\footnote{National Ocean and Atmospheric Administration, Boulder, Colorado, U.S.A.} and Christopher Heinz\footnote{Faculty of Mathematics and Economics, Ulm University, Germany.}}
\begin{document}

\date{}

\maketitle

\begin{abstract}
\noindent
Any decision making process that relies on a probabilistic forecast of future events necessarily requires a calibrated forecast. This paper proposes new methods for empirically assessing forecast calibration in a multivariate setting where the probabilistic forecast is given by an ensemble of equally probable forecast scenarios. Multivariate properties are mapped to a single dimension through a pre-rank function and the calibration is subsequently assessed visually through a histogram of the ranks of the observation's pre-ranks.  Average ranking assigns a pre-rank based on the average univariate rank while band depth ranking employs the concept of functional band depth where the centrality of the observation within the forecast ensemble is assessed.  Several simulation examples and a case study of temperature forecast trajectories at Berlin Tegel Airport in Germany demonstrate that both multivariate ranking methods can successfully detect various sources of miscalibration and scale efficiently to high dimensional settings. 

\vspace{3mm}
\end{abstract}

\section{Introduction}

Calibration, the statistical compatibility between a probabilistic forecast and the realized observation, is a fundamental property of any skillful forecast.  Formally, we say that the forecast is calibrated if, over the long run, events assigned a given probability are realized with the same empirical frequency.  Calibration is thus a critical requirement for optimal decision making and any decision aiding technique that relies on the forecast \citep{Lichtenstein&1977, Gneiting&2007}. 

In the case of  a univariate probabilistic forecast given by a continuous predictive distribution, \cite{Dawid1984} proposes the use of the probability integral transform (PIT) for calibration assessment.  That is, if $F$ is the cumulative distribution function (CDF) of a calibrated probabilistic forecast for the observation $y$, it holds that $F(y) \sim \mathcal{U}([0,1])$.  A randomized version of the PIT that applies to partly, or fully, discrete distributions is discussed in \cite{Czado&2009}.  For an ensemble of deterministic forecasts that approximate the predictive distribution, an equivalent tool is the rank of the observation $y$ in the forecast ensemble $x_1, \ldots, x_{m-1}$ \citep{Anderson1996, HamillColucci1997}.   The calibration of a large number of forecast cases may then be assessed empirically by plotting the histogram of the resulting PIT values or verification ranks \citep{Gneiting&2007}.  If the forecasts lack calibration, the shape of the PIT or the 
verification rank histogram may reveal the nature of the misspecification and 
thus provide a useful guidance to the improvement of the forecasting method.  For instance, a $\cup$-shaped histogram is an indication of underdispersion while a $\cap$-shape suggests overdispersion. 

To assess the calibration of multivariate ensemble forecasts, \cite{Gneiting&2008} propose a general two-step framework.  In the first step, the observation and the ensemble members are assigned univariate pre-ranks.  The rank of the observation is then given by the rank of its pre-rank. A multivariate calibration technique based on minimum spanning trees proposed by \cite{SmithHansen2004} and \cite{Wilks2004} seamlessly falls within this framework.  Alternatively, \cite{Gneiting&2008} propose a multivariate rank structure equal to that of the empirical copula.    A recent extension that applies to full distributions is given in \cite{ZiegelGneiting2013}.  While the multivariate rank histogram has been shown to work well for low-dimensional forecasts, see e.g. \cite{Schuhen&2012} and \cite{Moeller&2013}, the multivariate ordering in the first step seems to lack power in higher dimensions \citep{PinsonGirard2012}.  Alternative methods for high-dimensional calibration assessment have thus been called for \citep{Pinson2013, Schefzik&2013}. 
 
We propose two pre-ranking methods that complement the techniques of \cite{Gneiting&2008}, \cite{SmithHansen2004} and \cite{Wilks2004}.  The new methods are based on the concept of band depth for functional data introduced by \cite{Lopez-PintadoRomo2009} which relates to the graphical representation of the functional data curves.  That is, continuous or discrete curves are given a center-outward ordering according to the centrality of a curve within the collection of sample curves.  \cite{SunGenton2011, SunGenton2012} apply this concept to develop a box plot for the visualization and outlier-detection of functional data.  Viewing a discrete curve of length $d$ as a point in $d$-dimensional space, we define a pre-ranking method based on the band depth concept of \cite{Lopez-PintadoRomo2009}.  In the discrete case, the band depth essentially corresponds to the average centrality of the $d$ points.  As a second alternative, we thus also consider a pre-rank given by the average of the univariate ranks. 

The remainder of the paper is organized as follows.  In Section 2, we review the concept of band depth for discrete data and define the two multivariate ranking methods.  Section 3 and 4 provide the results of simulation studies where we investigate the influence of dimensionality and correlation, respectively, on the band depth ranks, the average ranks and the two previously proposed techniques.  A further comparison of the four techniques is provided in Section 4, where we assess the calibration of temporal trajectories of temperature forecasts over Germany.  The paper then ends with a discussion in Section 5. 

\section{Ranking multivariate data}\label{sec:definitions}

Let $S = \{ \bfx_1,\ldots \bfx_m\}$ denote a set of points in $\R^d$ or a $d$-dimensional subset thereof, with $\bfx_i = (x_{i1},\ldots,x_{id})$.  Here, we can think of $S$ as comprising an ensemble forecast with $m-1$ ensemble members and the corresponding observation $\mathbf{y} = \bfx_m$.  Following the general set-up of \cite{Gneiting&2008}, the rank of the observation in $S$ is calculated in two steps,
\begin{enumerate}
\item[(i)] apply a pre-rank function $\rho_S :  \R^d \rightarrow \R_+$ to calculate the pre-rank, $\rho_S(\bfx)$, of every $\bfx \in S$;
\item[(ii)] set the rank of the observation $\bfx_m$ equal to the rank of $\rho_S(\bfx_m)$ in $\{ \rho_S(\bfx_1),\ldots, \rho_S(\bfx_m)\}$ with ties resolved at random.  
\end{enumerate}
Under minimum spanning tree ranking, the pre-rank function $\rho_S^{\textup{mst}}(\bfx)$ is given by the length of the minimum spanning tree of the set $S \setminus \bfx$ \citep{SmithHansen2004, Wilks2004}.  Here, a spanning tree of the set $S \setminus \bfx$ is a collection of $m - 2$ edges such that all points in $S \setminus \bfx$ are used. The spanning tree with the smallest length is then the minimum spanning tree \citep{Kruskal1956}; it may e.g. be calculated using the {\tt R} package {\tt vegan} \citep{R2013}.  The multivariate ranking of \cite{Gneiting&2008}, on the other hand, is defined using the pre-rank function
\bl
\begin{equation}\label{eq:mrh}
\rho_S^{\textup m}(\bfx) = \sum\limits_{i=1}^{m}\mathbbm{1}\{\textbf{x}_i \preceq \textbf{x} \},
\end{equation}
\el
where $\mathbbm{1}$ denotes the indicator function and $\bfx_i \preceq \bfx$ if and only if $ x_{ik} \leq x_k $ for all $k=1,\ldots,d$.  \cite{Gneiting&2008} further consider an optional initial step in the ranking procedure in which the data is normalized in each component before the ranking. As the pre-rank functions proposed below are invariant to such pre-processing, we omit this step here.  

\subsection{Band depth rank}

\cite{Lopez-PintadoRomo2009} introduce a center-outward ordering of curves which they call band depth.  In the discrete case, it is defined as the proportion of coordinates of $\bfx \in S$ inside bands defined by subsets of $n$ points from $S$,
\bl
\begin{align}\label{eq:bd}
\textup{bd}_S^n(\bfx) = {m \choose n}^{-1} \frac{1}{d} \sum_{k=1}^d \sum_{1\leq i_1 < \ldots < i_n \leq m} & \mathbbm{1}\big\{ \min\{x_{i_1k},\ldots, x_{i_nk} \} \leq x_k \big\} \\
& \, \times \mathbbm{1} \big\{ x_k \leq \max \{ x_{i_1k}, \ldots, x_{i_nk}\}\big\}. \nonumber
\end{align} 
\el
Note that \cite{Lopez-PintadoRomo2009} refer to this version of the definition as modified band depth, in reference to the corresponding definition for continuous curves.  It holds that $0 \leq \textup{bd}_S^n(\bfx) \leq 1$ for all $\bfx \in S$ and it gets closer to $1$ the deeper, or more central, the point $\bfx$ is in the set $S$. Previous studies note that the resulting ordering of the elements in $S$ is robust to changes in the value of $n$ and we thus only consider the case $n=2$ which is equal to the simplical depth of \cite{Liu1990} and computationally very efficient \citep{Lopez-PintadoRomo2009, Sun&2013}.

From \eqref{eq:bd}, we obtain the band depth pre-rank function 
\bl
\begin{align}\label{eq:bd prerank}
\rho_S^{\textup{bd}} (\bfx) & = \frac{1}{d} \sum_{k=1}^d \sum_{1\leq i_1 < i_2 \leq m} \mathbbm{1}\big\{ \min\{x_{i_1k}, x_{i_2k} \} \leq x_k \leq \max \{ x_{i_1k}, x_{i_2k}\}\big\} \nonumber \\
& = \frac{1}{d} \sum_{k=1}^d \Big[ \textup{rank}_S(x_k) \big[ m - \textup{rank}_S(x_k)\big]  + \big[ \textup{rank}_S(x_k) - 1\big] \sum_{i=1}^m \mathbbm{1} \{ x_{ik} = x_k \}  \Big], 
\end{align}
\el
where $\textup{rank}_S(x_k) = \sum_{i = 1}^m \mathbbm{1} \{ x_{ik} \leq x_k\}$ denotes the rank of the $k$th coordinate of $\bfx$ in $S$.  If $x_{ik} \neq x_{jk}$ with probability $1$ for all $i,j \in \{1,\ldots m\}$ with $i \neq j$ and $k = 1,\ldots, d$, the band depth pre-rank function in \eqref{eq:bd prerank} further simplifies to 
\bl
\begin{equation}\label{eq:bd prerank simple}
\rho_S^{\textup{bd}} (\bfx) = \frac{1}{d} \sum_{k=1}^d \big[ m - \textup{rank}_S(x_k) \big] \big[ \textup{rank}_S(x_k) - 1 \big] + (m-1),
\end{equation}
\el
see also \cite{Sun&2013}. 

It is straightforward to see that the band depth rank of an observation $\mathbf{y}=\bfx_m$ is uniformly distributed if $\bfx_1,\ldots,\bfx_{m}$ are independent and identically distributed, which implies a calibrated ensemble forecast.  However, the interpretation of the resulting rank histogram is somewhat different than that of the classical univariate verification rank histogram.  As the example in Figure~\ref{fig:illustration}(a) shows, the band depth pre-rank assesses the centrality of the elements in $S$, with the most central element(s) attaining the highest rank(s) and the most outlying element(s) attaining the lowest rank(s).  A skew histogram with too many high ranks is thus an indication of an overdispersive ensemble while too many low ranks can result from either an underdispersive or biased ensemble.  As demonstrated in the simulation study in Section~\ref{sec:correlation}, a lack of correlation in the ensemble will result in a $\cup$-shaped histogram while an ensemble with too high correlations 
produces a $\cap$-shaped histogram.

\subsection{Average rank}

\begin{figure}[t]
\centering
\subfigure[Band depth ranking]{\includegraphics[width=0.45\textwidth]{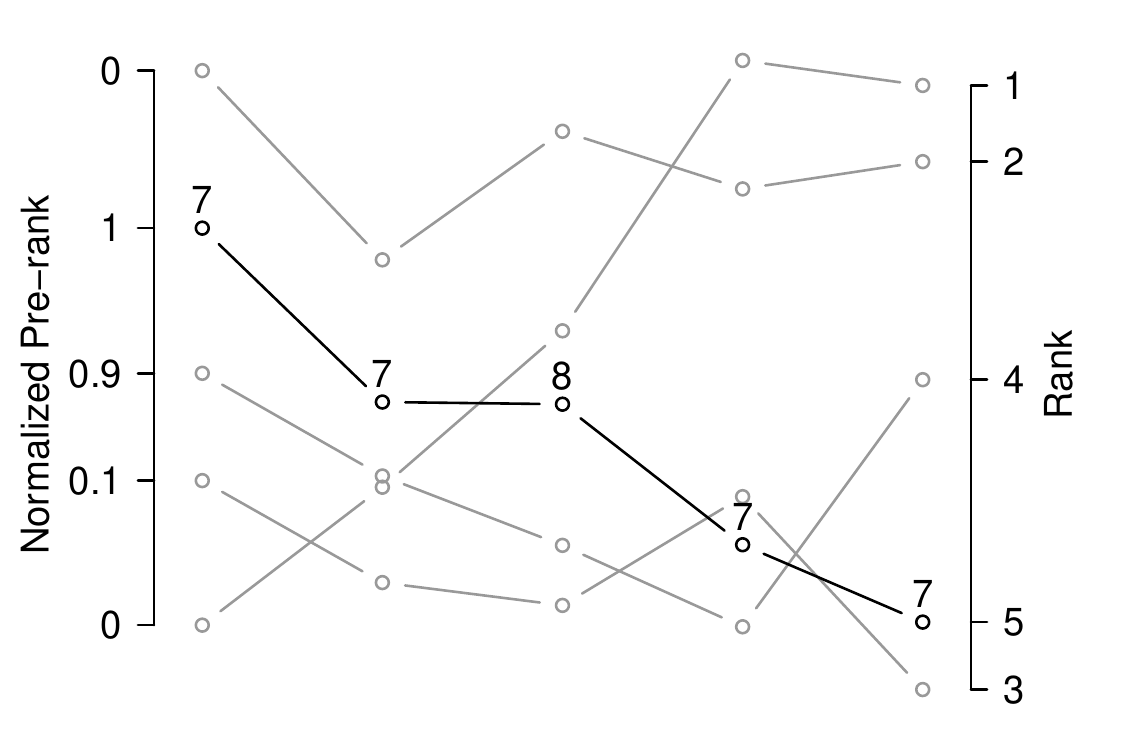}} \qquad
\subfigure[Average ranking]{\includegraphics[width=0.45\textwidth]{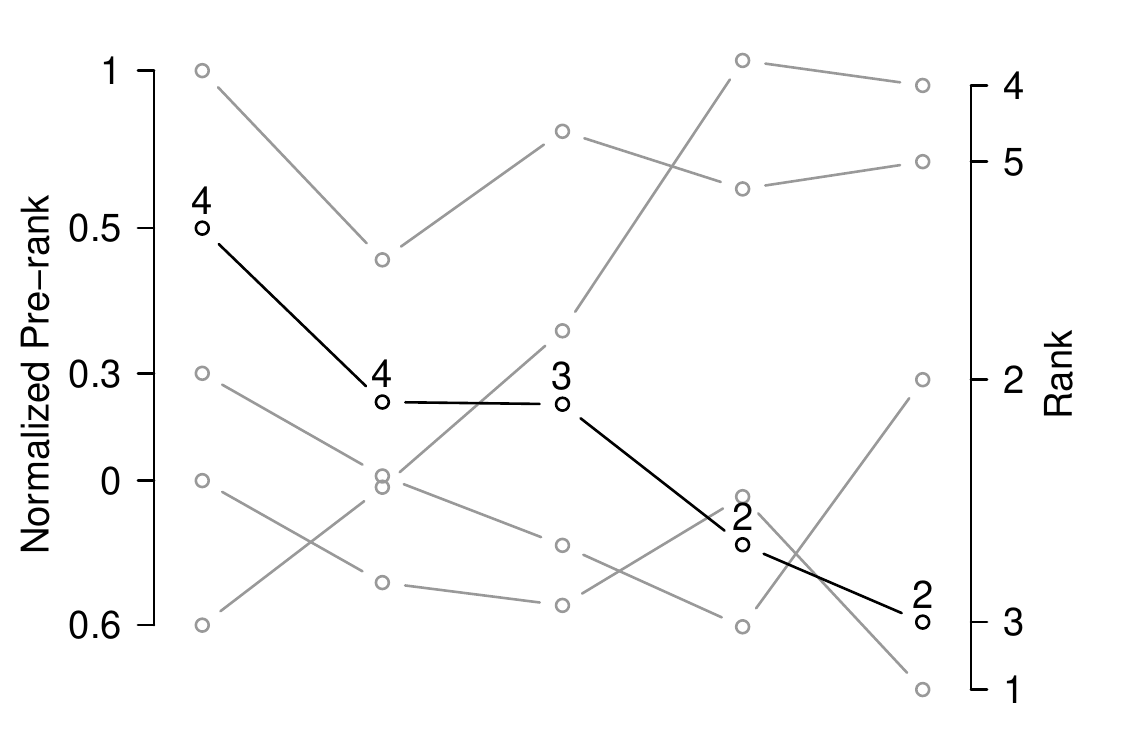}}
\caption{Illustration of (a) band depth, and (b) average pre-ranking for a multivariate temporal trajectory with $d=5$ time points. The normalized pre-ranks of each curve are given on the left and the resulting ranks on the right. The four ensemble forecast curves are indicated in gray and the observation curve in black. The numbers next to each point of the observation curve indicate the univariate pre-ranks.}\label{fig:illustration}
\end{figure}

The average rank is simply given by the average over the univariate ranks,
\begin{equation}\label{eq:average prerank}
\rho_S^{\textup{a}}(\bfx) = \frac{1}{d} \sum_{k=1}^d \textup{rank}_S(x_k). 
\end{equation}
An illustration of the average pre-ranking is given in Figure~\ref{fig:illustration}.  It follows directly from \eqref{eq:average prerank} that the resulting rank of the observation $\bfx_m$ in $S$ is uniform on $\{1,\ldots,m\}$ if the elements of $S$ are independent and identically distributed.  The average rank furthermore reduces to the classical univariate rank when $d=1$.  

The interpretation of the resulting histogram is similar to that of the univariate verification rank histogram. That is, if the forecasts are underdispersive the average rank histogram for the observation is $\cup$-shaped, an overdispersive ensemble results in a $\cap$-shaped histogram while a constant bias results in a triangular shaped histogram. As discussed in Section~\ref{sec:correlation} under- and overestimation of the correlation structure can furthermore result in over- and underdispersive histograms, respectively. 

\section{Histogram shape and the effect of dimensionality}

\begin{figure}[!hbpt]
\centering
\includegraphics[width=0.75\textwidth]{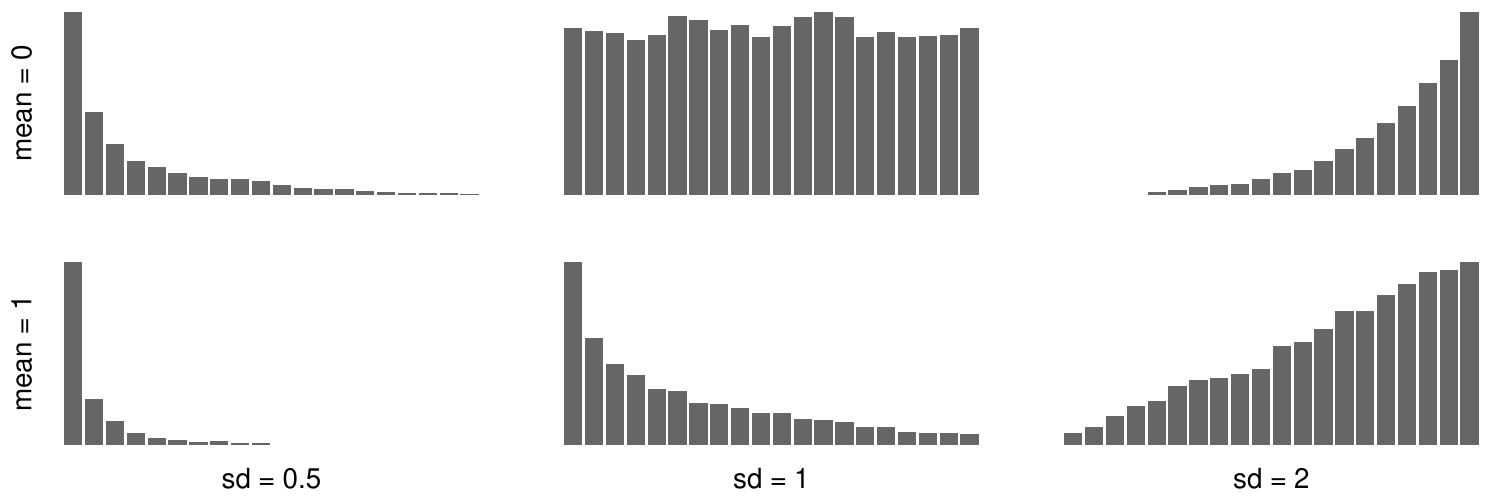}
\caption{Band depth rank histograms for observations in $d=3$ dimensions that follow independent standard Gaussian distributions while the $19$ ensemble members follow independent Gaussian distributions with parameters as indicated. The results are based on 10000 repetitions.}\label{fig:bd dim3}
\end{figure}

\begin{figure}[!hbpt]
\centering
\includegraphics[width=0.75\textwidth]{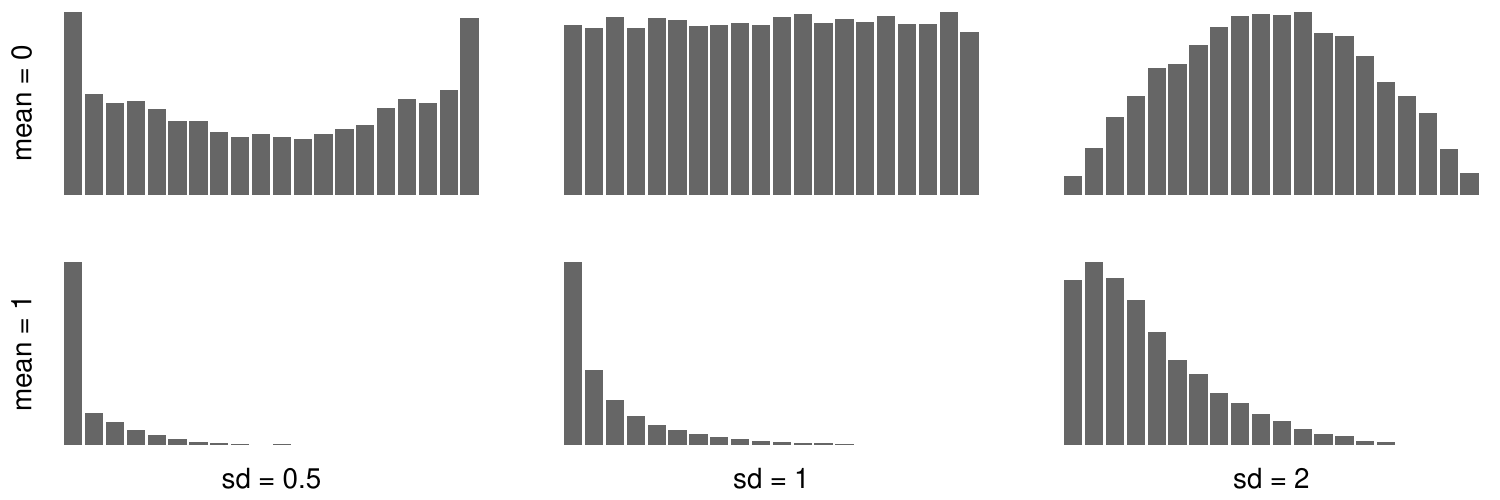}
\caption{Average rank histograms for observations in $d=3$ dimensions that follow independent standard Gaussian distributions while the $19$ ensemble members follow independent Gaussian distributions with parameters as indicated. The results are based on 10000 repetitions.}\label{fig:avg dim3}
\end{figure}

To demonstrate the shape of the histograms subject to over- and underdispersion as well as bias, we consider a simple simulation experiment where the observations follow an independent standard Gaussian distribution in each dimension.  Figure~\ref{fig:bd dim3} shows band depth rank histograms under this model in a low dimensional setting with $d=3$ and $m=20$.  The ensemble forecasts are also assumed to follow independent Gaussian distributions with mean $\mu \in \{0,1\}$ and standard deviation $\sigma \in \{0.5, 1, 2\}$.  When the forecasts  are underdispersive or have a constant bias, the observation curve is often among the most outlying curves resulting in too many low ranks. Similarly, if the forecasts are overdispersive, the observation curves are too central on average, resulting in too many high ranks.  Figure~\ref{fig:avg dim3} shows the average rank histograms for the same setting.  Here, the interpretation of the average ranks is equivalent to that of the standard univariate rank histogram.  The histogram shape clearly indicates overdispersion in the forecast through a $\cap$-shape, underdispersion through a $\cup$-shape and bias via a skew, triangular shaped histogram. 

\begin{figure}[!hbpt]
\centering
\includegraphics[width=\textwidth]{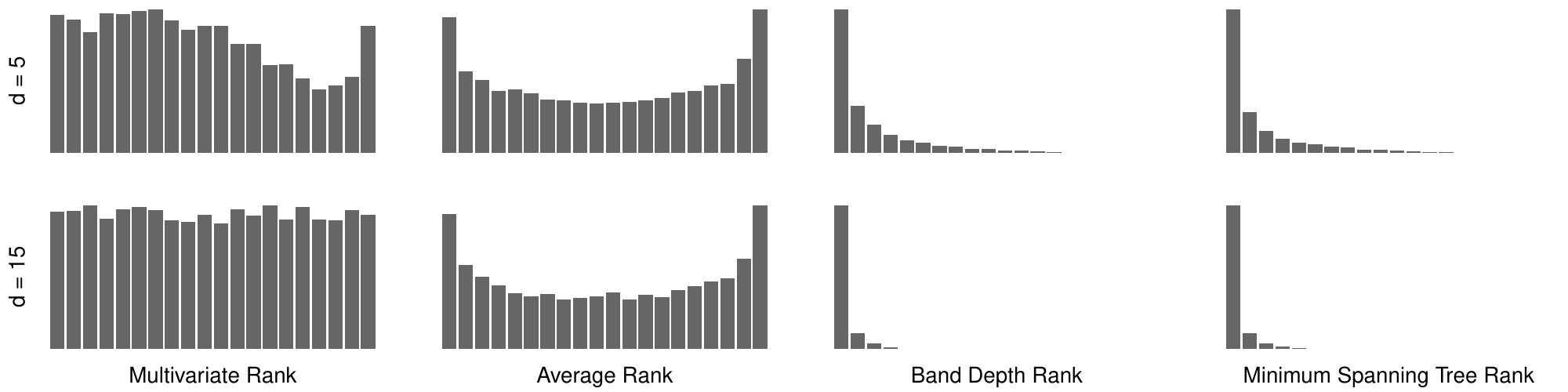}
\caption{Multivariate ranking of observations in dimension $d=5$ (top row) and $d=15$ (bottom row) that follow independent standard Gaussian distributions when the 19 ensemble member forecasts are underdispersed following independent zero-mean Gaussian distributions with standard deviation of 0.5.}\label{fig:sd 0.5}
\end{figure}

\begin{figure}[!hbpt]
\centering
\includegraphics[width=\textwidth]{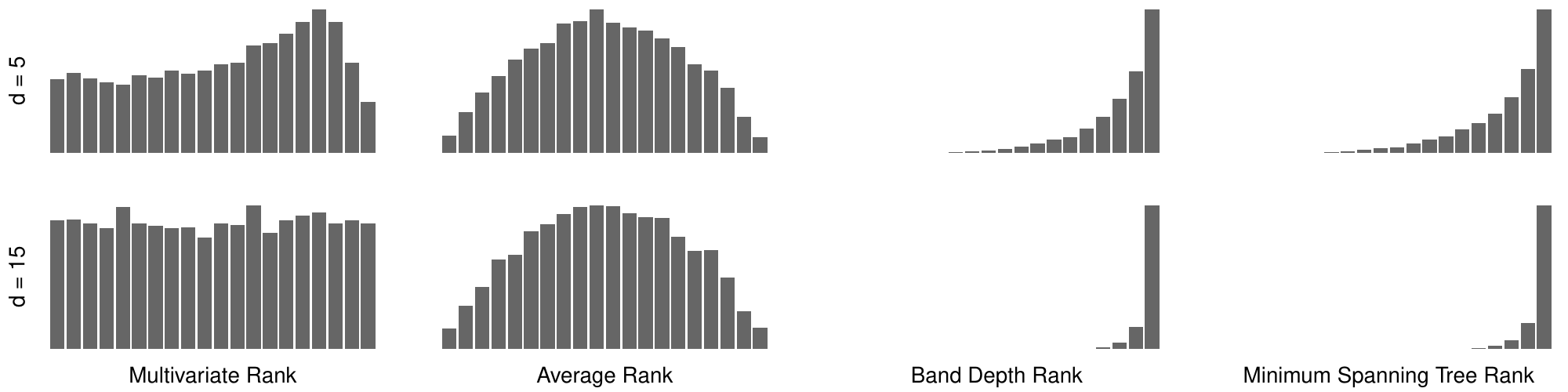}
\caption{Multivariate ranking of observations in dimension $d=5$ (top row) and $d=15$ (bottom row) that follow independent standard Gaussian distributions when the 19 ensemble member forecasts are overdispersed following independent zero-mean Gaussian distributions with standard deviation of 2.}\label{fig:sd 2}
\end{figure}

Figure~\ref{fig:sd 0.5} and \ref{fig:sd 2} demonstrate the effect of increasing dimensionality on the four multivariate ranking methods discussed in Section~\ref{sec:definitions} subject to under- and overdispersion, respectively. While we still assume the ensemble consists of 19 members, the dimensionality of the data is here increased to $5$ and $15$ dimensions.  This setting may seen somewhat extreme in that we attempt to represent the multivariate correlation structure in $15$ dimensions with only $19$ trajectories.  However, this is common e.g.\ in atmospheric sciences, where due to computational limitations ensembles of similar magnitude are used to represent very high dimensional multivariate distributions. 

The average rank histograms for both examples appear unchanged compared to the low dimensional example in Figure~\ref{fig:avg dim3} while for the band depth rank, the evidence of miscalibration seem to get stronger with higher dimensions.  The minimum spanning tree ranking provides a center-outward ordering of the curves similar to statistical depth functions \citep{Gneiting&2008, ZuoSerfling2000} and for the examples here, the shape of the minimum spanning tree rank histograms is nearly identical to that of the band depth rank histograms.  As reported in \cite{PinsonGirard2012}, we observe identifiability issues with the multivariate ranking of \cite{Gneiting&2008} in higher dimensions.  In $5$ dimensions, only the upper half of the ranks indicates miscalibration and the multivariate rank histograms appear close to uniform when $d=15$ even though the forecasts are severely miscalibrated. The reason for this can be seen by considering the example in Figure~\ref{fig:illustration}, where, due to crossing of 
the curves, four out of the five curves would obtain a multivariate pre-rank of 1.  

Addtional simulation studies show that miscalibration is generally easier to detect in larger ensembles than in small ensembles (results not shown). While these results holds across the different pre-ranking techniques, it appears that the curse of dimensionality observed for the multivariate ranking in Figures~\ref{fig:sd 0.5} and \ref{fig:sd 2} cannot be avoided by increasing the size of the forecast ensemble.  

\section{Assessing deviations in the correlation structure}\label{sec:correlation}

An appropriate modeling of the correlation between the different components is an important aspect of multivariate predictions.  It is not entirely obvious from their definition why the band depth and the average rankings are sensitive to misspecification of the correlation structure.  This can be demonstrated by comparing the variances of the pre-ranks under different dependence strengths.  First, consider the extreme case where the observations are fully dependent (i.e.\ identical) and the forecasts are independent across the different components.  Assuming, as before, that the different curves are pairwise independent, the rank of the $i$th random curve $\mathbf{X}_i$ is uniformly distributed on $\{1,\ldots,m\}$ for each component $k = 1,\ldots,d$.  Under the pre-rank functions in \eqref{eq:bd prerank simple} and \eqref{eq:average prerank} it follows that 
\begin{equation}\label{eq:expected prerank}
\E \big( \rho_S^{\textup{a}} (\mathbf{X}_i) \big) = \frac{m + 1}{2}, \quad \E \big( \rho_S^{\textup{bd}} (\mathbf{X}_i)\big) = \frac{m^2+3m - 4}{6}, \quad i = 1, \ldots, m.   
\end{equation}

For simplicity, we assume that the number $m-1$ of forecast curves is high enough, so that we can neglect the different dependence structure of the observation curve when calculating the variance of the pre-rank function for the forecast curves. For the average ranking we obtain 
\begin{align}
 \Var\big( \rho_S^{\textup{a}} (\mathbf{X}_i) \big) & \approx \frac{m^2-1}{12d}, & i=1,\ldots,m-1, \label{eq:var avg forecast}\\
 \Var\big( \rho_S^{\textup{a}} (\mathbf{X}_i) \big) & = \frac{m^2-1}{12d} + \frac{(m-1)^2 (d-1)}{12d}, & i = m, \label{eq:var avg obs}
\end{align}
while the band depth ranking results in 
\begin{align}
\Var \big( \rho_S^{\textup{bd}} (\mathbf{X}_i) \big) & \approx \frac{(m+1)(m-1) (7m^2 + 8m + 12)}{60 d}, & i=1,\ldots,m-1, \label{eq:var bd forecast}\\
\Var \big( \rho_S^{\textup{bd}} (\mathbf{X}_i) \big) & = \frac{(m+1)(m-1) (7m^2 + 8m + 12)}{60 d} \nonumber \\
& \quad + \frac{(m^4 - 6m^3 + 13m^2 - 12m + 4)(d-1)}{180d}, & i = m. \label{eq:var bd obs}
\end{align}
Details of the derivations are given in the appendix. 

That is, the variance of the pre-rank for the observation curve (which was assumed constant over all components) is much larger than that of the forecasts curves (which were assumed independent across all components) for both pre-rank functions. It is thus more likely that we observe a very low or a very high pre-rank for the observation than for each ensemble member forecast which again leads to proportionally larger number of low and high ranks for the observation resulting in a $\cup$-shaped histogram. 

\subsection{Gaussian autoregressive processes}\label{sec:AR}

We now consider an example where $\mathbf{y} \in \R^d$ is a temporal trajectory of a real valued variable observed at $d$ equidistant time points $t=1,\ldots,d$.  That is, the observation is a realization of a zero-mean Gaussian AR(1) (autoregressive) process $\mathbf{Y}$ with 
\begin{equation}\label{eq:AR1}
\textup{Cov} (Y_i,Y_j) = \exp (- |i - j|/\tau), \quad \tau > 0. 
\end{equation}
The process $\mathbf{Y}$ thus has standard Gaussian marginal distributions while the parameter $\tau$ controls how fast correlations decay with time lag. We set $\tau=3$ for $\mathbf{Y}$ and consider ensemble forecasts of the same type but with a different parameter value $\tau$.  It follows from this construction that a univariate calibration test at a fixed time point would not detect any miscalibration in the forecasts. 

\begin{figure}[!hbpt]
\centering
 \includegraphics[width=\textwidth]{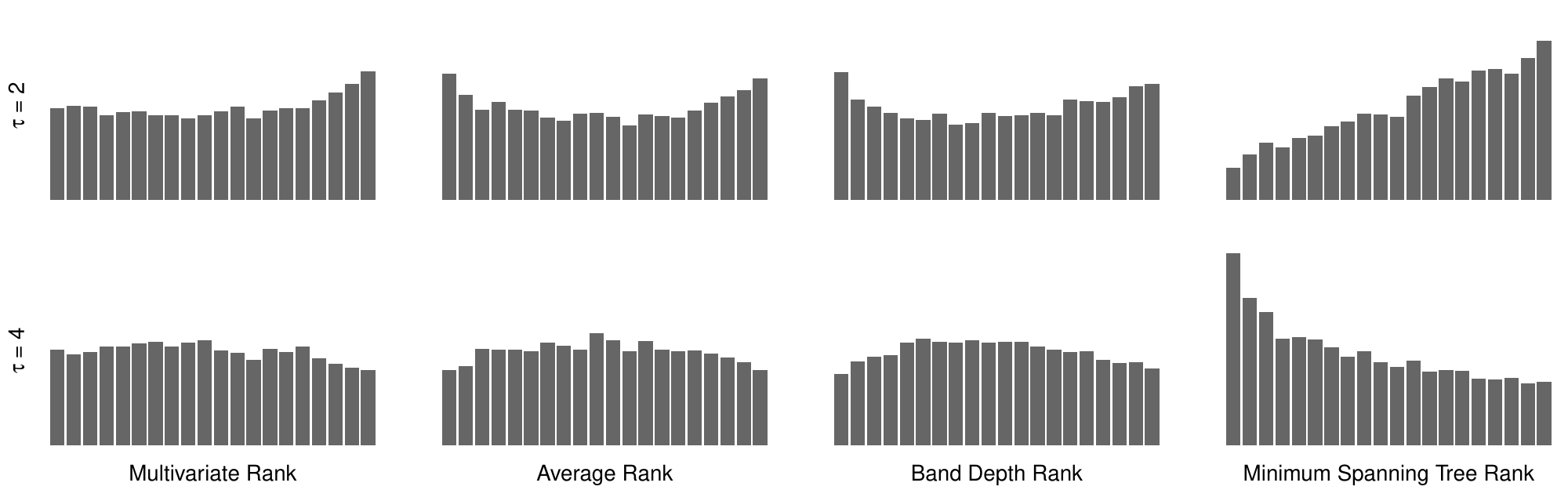}
 \caption{Simulation study to compare the sensitivity of the multivariate rank histogram, the band depth rank histogram and the average rank histogram to misspecification of the dependence structure. The observations follow an AR(1) process at time $t = 1,\ldots,5$ with the dependence structure given in \eqref{eq:AR1} for $\tau=3$ while the ensemble forecasts follow the same model with $\tau=2$ (top row) and $\tau=4$ (bottom row). The results are based on 10000 repetitions with 19 ensemble members in each iteration.}
 \label{fig:AR1-d5}
\end{figure}

Rank histograms for $d=5$ and $m=20$ where the forecast model has either $\tau=2$ or $\tau=4$ are shown in Figure~\ref{fig:AR1-d5}.  While all four calibration assessment methods are able to detect the miscalibration, the multivariate rank histogram suffers from identifiability issues with many low and identical pre-ranks resulting in a flattening out of the left side of the histograms.   The band depth and the average rankings, on the other hand, seem quite sensitive to the model misspecification resulting in $\cup$-shape histograms when the correlations decay too fast in the forecasts and $\cap$-shaped histograms in the opposite situation.  Here, the minimum spanning tree histogram gives the clearest indication of miscalibration.

Tables~\ref{tab:AR1mean} and \ref{tab:AR1variance} demonstrate the effect of dimensionality and ensemble size on the average and band depth rank histograms in Figure~\ref{fig:AR1-d5}. That is, we report the mean rank and the rank variance for both the observation and a randomly selected ensemble member under the two ranking methods when the observation follows the model in \eqref{eq:AR1} with $\tau=3$ while $\tau=2$ for the forecasts.  This example is similar to the example at the beginning of this section which can be considered the extreme case with $\tau = \infty$ for the observation and $\tau = 0$ for the forecast. 

\begin{table}[!hbpt]
\centering
\caption{Mean ranks over 30000 repetitions for average ranking and band depth ranking under a zero-mean Gaussian AR(1) model with the exponential covariance function in \eqref{eq:AR1} with $\tau = 3$ for the observation and $\tau =2$ for the forecasts. }\label{tab:AR1mean}
\begin{tabular}{lcccccccc}
\toprule
& \multicolumn{4}{c}{Average} & \multicolumn{4}{c}{Band depth} \\  
\cmidrule(r){2-5} \cmidrule(l){6-9}
& {\small $m=20$} & {\small $m=100$} & {\small $m=200$} & {\small $m=500$} & {\small $m=20$} & {\small $m=100$} & {\small $m=200$} & {\small $m=500$} \\
\midrule 
\multicolumn{4}{l}{Observation} \\ 
\midrule 
$d=5$ & 10.5 & 50.4 & 100.0 & 251.5 & 10.7 & 51.7 & 102.2 & 256.8 \\
$d=100$  & 10.6 & 50.4 & 101.0 & 250.8 & 10.6 & 50.8 & 101.7 & 253.2   \\
$d=200$ & 10.5 & 50.4 & 100.2 & 251.2 & 10.5 & 50.9 & 101.8 & 251.5 \\
$d=500$ & 10.5 & 50.7 & 100.3 & 249.7 & 10.5 & 50.9 & 100.9 & 251.4 \\
\midrule
\multicolumn{9}{l}{Randomly selected ensemble member} \\
\midrule 
$d=5$ & 10.5 & 50.7 & 100.4 & 249.5 & 10.5 & 50.6 & 100.6 & 248.6 \\
$d=100$ & 10.5 & 50.7 & 101.3 & 250.7 & 10.5 & 50.2 & 100.5 & 251.1 \\
$d=200$ & 10.5 & 50.3 & 100.4 & 250.7 & 10.5 & 50.3 & 100.5 & 252.3 \\
$d=500$ & 10.5 & 50.3 & 100.4 & 250.6 & 10.5 & 50.5 & 100.4 & 251.2 \\
\bottomrule
\end{tabular}
\end{table}

\begin{table}[!hbpt]
\centering
\caption{Rank variance over 30000 repetitions for average ranking and band depth ranking under a zero-mean Gaussian AR(1) model with the exponential covariance function in \eqref{eq:AR1} with $\tau = 3$ for the observation and $\tau =2$ for the forecasts. }\label{tab:AR1variance}
\begin{tabular}{lcccccccc}
\toprule
& \multicolumn{4}{c}{Average} & \multicolumn{4}{c}{Band depth} \\  
\cmidrule(r){2-5} \cmidrule(l){6-9}
& {\small $m=20$} & {\small $m=100$} & {\small $m=200$} & {\small $m=500$} & {\small $m=20$} & {\small $m=100$} & {\small $m=200$} & {\small $m=500$} \\
\midrule 
\multicolumn{4}{l}{Observation} \\ 
\midrule 
$d=5$ & 37 & \,\,\,940 & 3773 & 23428 & 37 & 946 & 3749 & 23690 \\
$d=100$  & 40 & 1004 & 4042 & 25431 & 38 & 989 & 3982 & 24604   \\
$d=200$ & 39 & 1006 & 4002 & 25524 & 38 & 984 & 3949 & 24747 \\
$d=500$ & 39 & 1014 & 4052 & 25629 & 38 & 992 & 3965 & 24891 \\
\midrule
\multicolumn{9}{l}{Randomly selected ensemble member} \\
\midrule 
$d=5$ & 33 & \,\,\,830 & 3319 & 20849 & 33 & 835 & 3341 & 20891 \\
$d=100$ & 33 & \,\,\,837 & 3323 & 20663 & 33 & 825 & 3331 & 20715 \\
$d=200$ & 33 & \,\,\,828 & 3316 & 21008 & 33 & 833 & 3315 & 20920 \\
$d=500$ & 33 & \,\,\,833 & 3320 & 20763 & 33 & 835 & 3336 & 20825 \\
\bottomrule
\end{tabular}
\end{table}

In the current example, dimensionality has only a minimal effect on the results while the size of the ensemble substantially affects the resulting values due to the varying number of possible ranks. As the serial dependence of the forecasts is too weak, the forecast ranks concentrate more strongly around the mean than the obseration ranks resulting in $\cup$-shaped histograms as those displayed in the top row of Figure~\ref{fig:AR1-d5}.  This difference in the rank variance appears to be somewhat stronger for the average ranking than for the band depth ranking.  For the band depth ranking, we moreover observe a slight shift of the mean rank.  This follows from the fact that the distribution of the band depth rank, a quadratic function of the univariate ranks, is slightly skewed such that difference in the variance of the pre-ranks may cause differences in the mean rank. 

When the forecast model has the parameter value $\tau = 4$ as displayed in the bottom row of Figure~\ref{fig:AR1-d5}, we observe similar effects of dimensionality and ensemble size as those reported in Tables~\ref{tab:AR1mean} and \ref{tab:AR1variance}. However, as this example has too strong serial dependence in the forecasts, the rank variance of the observations is here lower than that of the forecasts (results not shown).   

\subsection{Autoregressive vs. more complex correlation functions}\label{sec:4.2}

\begin{figure}
 \includegraphics[width=\textwidth]{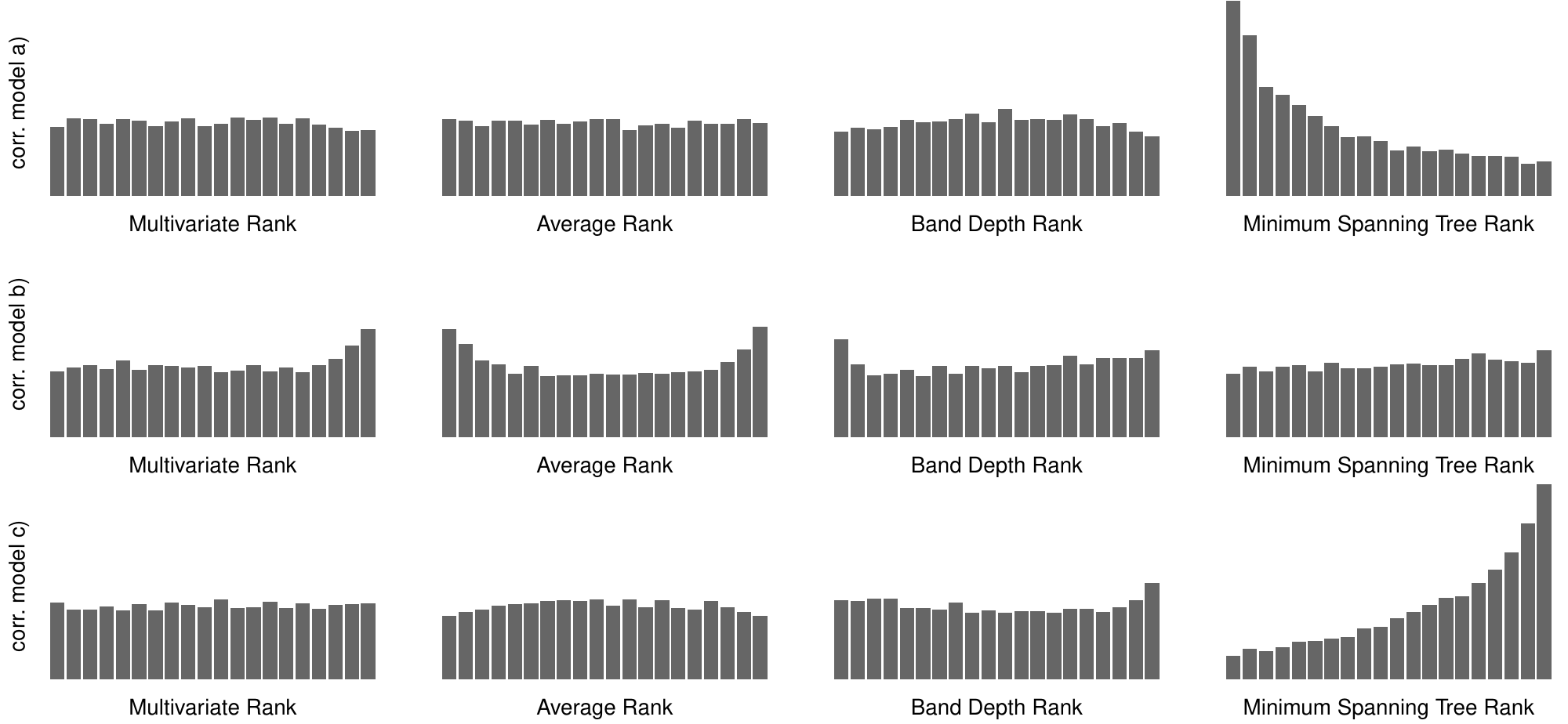}
 \caption{Simulation study to compare the sensitivity of the four multivariate ranking methods to miscalibration in the dependence structure. The observations follow the correlation models a), b), or c) (from top to bottom) at time $t = 1, \ldots, 15$  while the forecasts follow an AR(1) process with scale parameter $\tau=3$. The results are based on 10000 repetitions with an ensemble of size 19.}
 \label{FigAR1andOthers}
\end{figure}

Here, we consider Gaussian processes on $t=1,\ldots,d$ where the observation follows the model in \eqref{eq:AR1} with $\tau=3$ while the components of the observation curve have a more complex correlation structure. That is, we consider the correlation models 
\begin{enumerate}
 \item[a)] $\textup{Cov} (Y_i,Y_j) = \exp(-|i-j|/4.5) \big(0.75+0.25\cos(\pi|i-j|/2)\big)$
 \item[b)] $\textup{Cov} (Y_i,Y_j) = \big(1+|i-j|/2.5\big)^{-1}$
 \item[c)] $\textup{Cov} (Y_i,Y_j) = \mathbbm{1}\big\{ |i-j| \leq 5 \big\} \big(1-|i-j|/5\big)$
\end{enumerate}
Correlation function a) is a damped cosine that oscillates around the exponential model (\ref{eq:AR1}) with $\tau=3$. The correlation functions b) and c) differ from this exponential model in that they have much stronger correlations at larger time lags, or zero correlations for larger time lags, respectively. 

Figure~\ref{FigAR1andOthers} shows the resulting histograms for $d=15$ and $m=20$.  When the observations follow correlation model a), the univariate ranks cancel out by averaging which results in a flat average rank histogram, while the minimum spanning tree histogram detects the false correlation structure very well and the band depth rank histogram also indicates miscalibration. For the long range dependence model the opposite situation occurs in that the average rank histogram gives the clearest indication of miscalibration while the minimum spanning tree histogram is almost flat. 

The last model c) with zero correlations beyond lag $5$ finally presents a situation where the average rank and band depth rank histograms behave in the opposite way, the former being slightly $\cap$-shaped and the latter being slightly $\cup$-shaped. This suggests that the average rank histogram is more strongly affected by correlations at larger lags (which are overpredicted here) while the band depth rank histogram and the minimum spanning tree histogram are more sensitive to misspecifications of correlations at short lags (which are underpredicted here).

\section{Calibration of temperature forecast trajectories}\label{sec:ex}

We illustrate the use of the multivariate verification tools discussed above in the setting of probabilistic weather forecasting, where ensembles of weather predictions for the same location, time and weather variable are generated in order to represent forecast uncertainty \citep{Palmer2002, GneitingRaftery2005, Schefzik&2013}. Specifically, we consider ensemble temperature forecasts at Berlin Tegel issued by the ensemble prediction system (EPS) of the European center for medium-range weather forecasts (ECMWF) with lead times of 6h, 12h, ..., 72h \citep{Molteni&1996, LeutbecherPalmer2008}. The EPS is initialized at 0000 UTC, consists of 50 ensemble members, and will be evaluated during the period from October 10, 2010 to December 31, 2012 using observational data from the local meteorological station as the truth. The ECMWF forecasts used here are freely available from the TIGGE repository at \url{http://apps.ecmwf.int/datasets/data/tigge/}. 

The univariate rank histograms (not shown here) suggest that these raw ensemble forecasts have a systematic under forecasting bias at Berlin Tegel and are underdispersive at all considered lead times.  We use a simple post-processing method to remove bias and adjust the ensemble spread for each lead time separately.  Denoting by $\bar{\mathbf x}$ the mean of the 50 ensemble members (this is a vector with 12 components, one for each lead time) we obtain a bias-corrected mean $\mu$ by fitting a linear regression model $\mu_i=a_i+b_i\bar{x}_i$, separately for each component, to the corresponding observations $y_i$. For each forecast day the preceding $50$ days are taken as training data so that we always have $50$ forecast-observation pairs to fit the regression model. This is a compromise between flexible adaptation to seasonal changes on the one hand and gathering sufficient data to permit stable model fitting on the other hand, see e.g. \cite{Gneiting&2005} and \cite{Raftery&2005}. 

\begin{figure}
 \includegraphics[width=\textwidth]{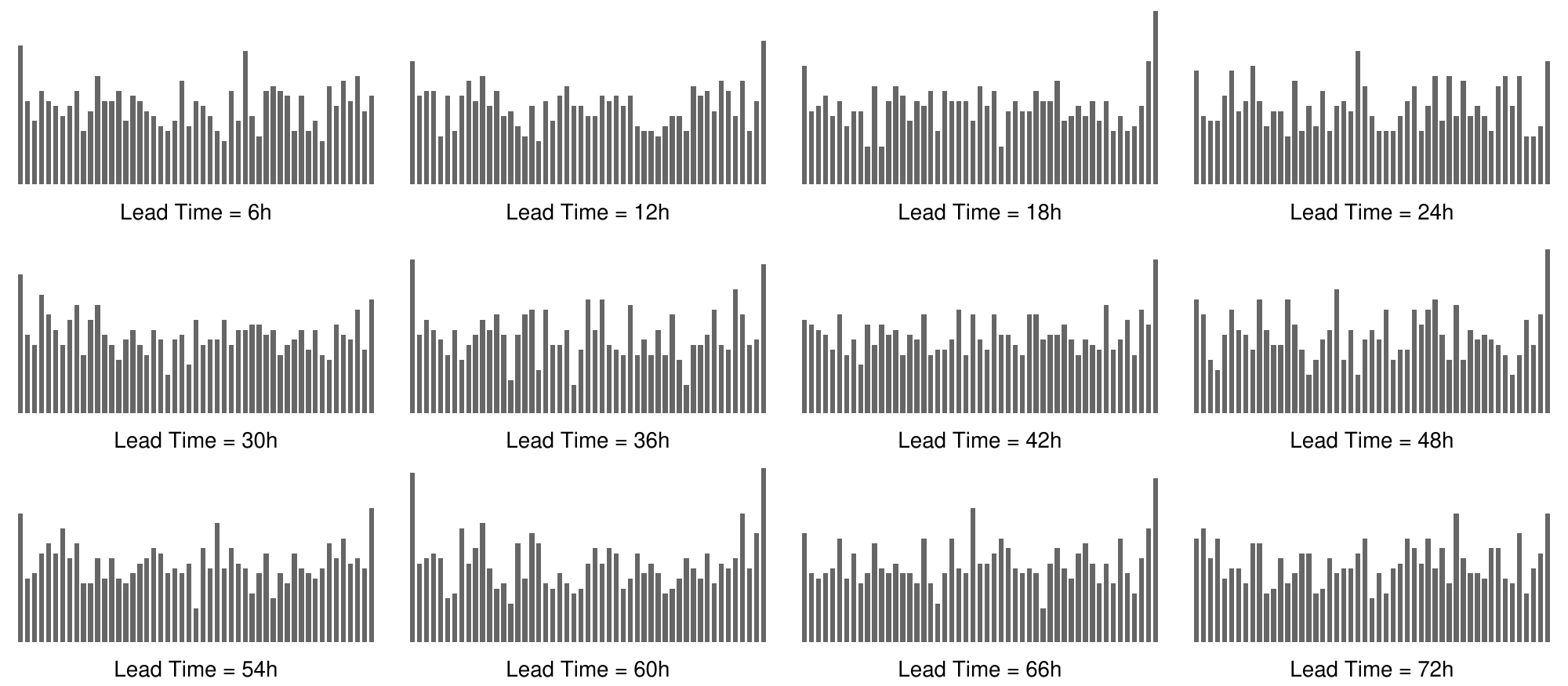}
 \caption{Univariate rank histogram of the bias-corrected error dressing forecasts for lead times 6h, 12h, ..., 72h at Berlin Tegel, each of them based on 823 verification days.}
 \label{FigErrorDressingUnivariate}
\end{figure}

To adjust the ensemble spread, we use the ``error dressing''  approach of \cite{RoulstonSmith2003}, building a new ensemble by sampling from the errors $\varepsilon_{ij}=y_{ij}-\mu_{ij}$ of the bias-corrected forecasts on the respective training days $j=1,\ldots,50$ for lead time $i = 1,\ldots,12$. To create an ensemble that appropriately represents the prediction uncertainty we additionally inflate $\varepsilon_{ij}$ to adjust for the uncertainty in the bias correction \citep[Section~3.5]{Faraway2004}. The ensemble obtained in this way is unbiased and nearly calibrated for individual lead times, see Figure~\ref{FigErrorDressingUnivariate}.  

We then consider three different strategies to model dependencies of forecast errors at different lead times,

\begin{enumerate}
 \item[(i)] ignore multivariate dependencies and perform the error dressing separately for each lead time;
 \item[(ii)] perform the error dressing separately for each lead time but use empirical copula coupling \cite[ECC,][]{Schefzik&2013} in a second step to transfer the dependence structure from the raw ECMWF ensemble to the error dressing ensemble;
 \item[(iii)] draw the errors from a zero-mean multivariate normal distribution with the empirical covariance matrix of the forecast errors over all lead times, where the variance is inflated as suggested above. 
\end{enumerate}

\begin{figure}[t]
\centering
\includegraphics[width=\textwidth]{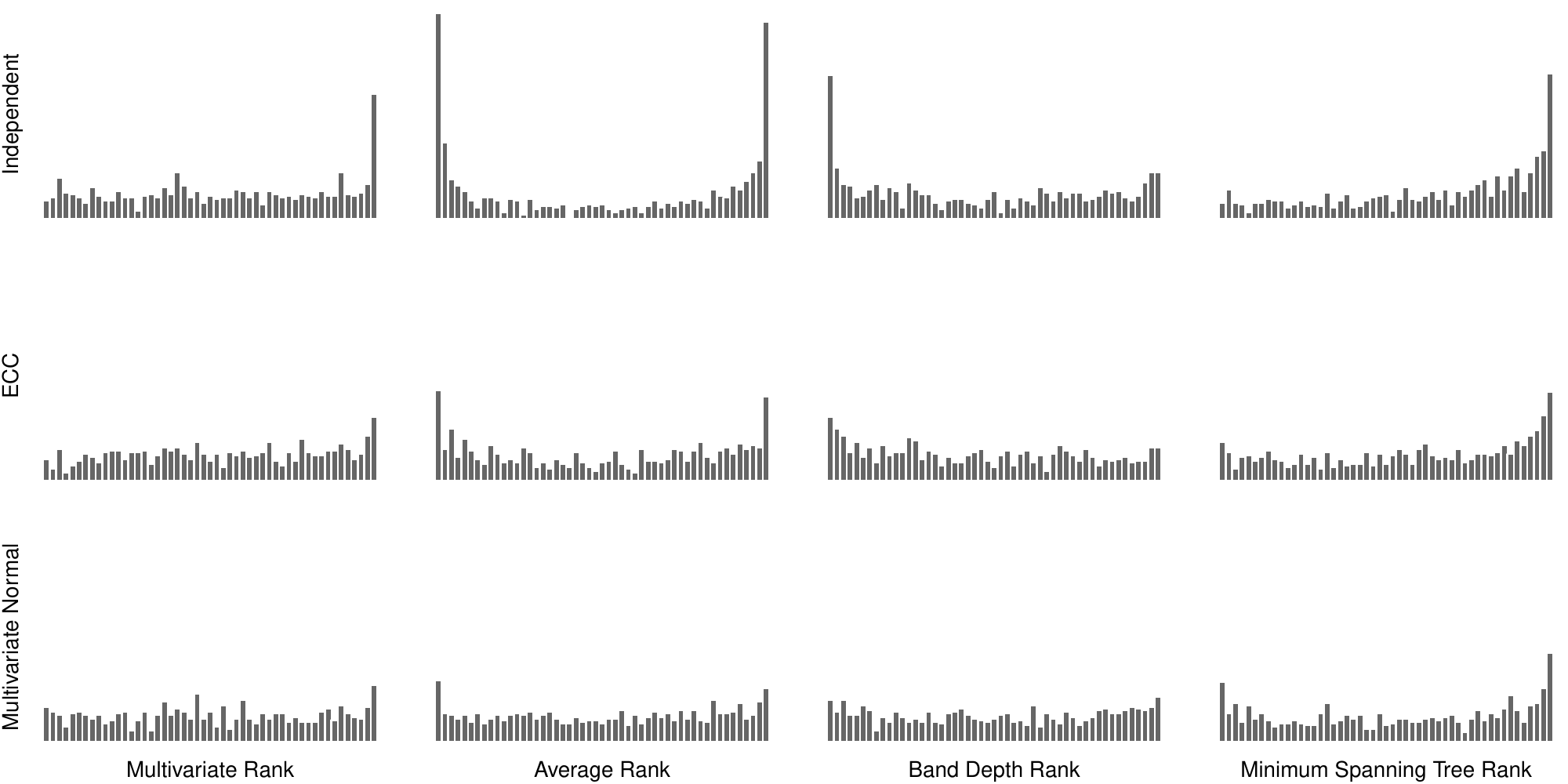}
\caption{Multivariate rank histograms (left), band depth rank histograms (middle) and average rank histograms (right) of the bias-corrected error dressing forecasts with independent error sampling (top), under ECC (middle) and with multivariate normal error sampling (bottom).  The results are based on forecasts for 12 lead times on 823 verification days at Berlin Tegel. }\label{FigErrorDressingMultivariate}
\end{figure}

While all three strategies result in similar marginal distributions, the multivariate calibration assessment in Figure~\ref{FigErrorDressingMultivariate} reveals substantial differences.  When the statistical postprocessing is performed independently for each lead time, the average rank histogram exhibit a $\cup$-shape indicating a lack of correlation between lead times in the forecasts.  The band depth rank histogram is skew towards the lowest ranks indicating that the forecasts are too outlying on average and both the minimum spanning tree and the multivariate rank histograms are skewed towards the higher ranks.  However, as the average rank histogram is symmetric, we would expect the outlying observation curves to have both too low ranks as well as too high ranks on average. We thus observe here a flattening out of the lower ranks in the multivariate rank histogram due to degeneracy in the pre-ranking; on any given day, at least half the curves are assigned a multivariate pre-rank of $1$.

The ECC multivariate postprocessing of \cite{Schefzik&2013} significantly improves the calibration of the independent postprocessing, though the observation curves are still somewhat too outlying.  For the multivariate normal error sampling, the histograms appear quite close to uniform with a minor divergence towards a $\cup$-shape in both the minimum spanning tree rank histogram and the average rank histogram. An alternative forth multivariate postprocessing option is to apply univariate normal error models followed by ECC. This option leads to calibration results nearly identical to the current results for ECC.    

\section{Discussion}

In this paper, we propose two new methods for assessing the calibration of multivariate forecasts where the predictive distribution is represented by a forecast ensemble.  Band depth ranking is based on the concept of band depth for functional data, originally proposed by \cite{Lopez-PintadoRomo2009} and previously employed to create box plots for functional data \citep{SunGenton2011, SunGenton2012, Sun&2013}. The somewhat simpler alternative, average ranking, employs the average over the univariate ranks.  As demonstrated in several simulated and real data examples, both methods seem to correctly identify various sources of miscalibration in the forecast. Furthermore, they escape the curse of dimensionality affecting the multivariate ranking of \cite{Gneiting&2008} as e.g. discussed by \cite{PinsonGirard2012}. The minimum spanning tree ranking of \cite{SmithHansen2004} and \cite{Wilks2004} can be more sensitive to misspecifications than the new methods proposed here. However, the resulting histograms seem to provide less information on the type of misspecification.  

The band depth concept of \cite{Lopez-PintadoRomo2009} is but one of a multitude of statistical depth functions for multivariate data that provide a center-outward ordering of the data \citep{ZuoSerfling2000}.  While we have here chosen the band depth due to its computational efficiency and interpretability of the resulting histograms, other depth functions might be equally appropriate for this purpose.  As the band depth ranking assesses the centrality of the observation within the forecast ensemble, the sign of a potential bias cannot be learned from the shape of the histogram.  Average ranking, on the other hand, distinguishes between positive and negative bias and effects where the forecasts exhibit a positive bias in a subset of the dimensions and a negative bias in a different subset might cancel out.  Such effects can, however, easily be detected through univariate calibration assessment in each dimension. 

Our examples, in particular the examples in Section~\ref{sec:4.2}, suggest that there is no single best pre-ranking method as all the methods may fail in detecting miscalibration. These methods project the multivariate quantity on a different univariate aspect and, in the process, lose information on other aspects.  Our overall recommendation is thus to study histograms of different type before drawing conclusions.  Furthermore, multivariate techniques should first and foremost complement univariate methods by effectively detecting features of miscalibration that cannot be found by studying the marginal distributions only. Conversely, ensuring marginal calibration in a first step can rule out the possibility of some compensating effects e.g.\ of marginal variances and correlations between different components.

Multivariate ranks relate to the multi-dimensional Smirnov two sample test proposed by \cite{Bickel1969}.  Formal tests of uniformity can also be applied to the resulting ranks and this has been studied by several authors for univariate PIT or rank histograms, see e.g. \cite{Gneiting&2007} and references therein.  However, as dicussed by both \cite{Hamill2001} and \cite{Gneiting&2007}, the use of formal tests is often complicated by the intricate dependence structures between the individual forecast cases. This holds, in particular, for partially overlapping forecast trajectories as discussed in Section~\ref{sec:ex} or spatially aggregated forecasts. 

Although calibration is an essential feature of a skillful forecast, a general forecast verification framework should consider a number of different aspects. \cite{Gneiting&2007} state that the goal of probabilistic forecasting is to ``maximize the sharpness with respect to calibration''.  That is, given a group of forecasts that all appear close to calibrated, we should choose the forecast with the highest information content. For predictive distributions or forecast ensembles, this can be attained by choosing the forecast with the smallest spread. More generally, proper scoring rules offer a verification framework under which various aspects of the forecast can be assessed, including calibration and sharpness. A comprehensive review of proper scoring rules is given in \cite{GneitingRaftery2007}. 

\section*{Acknowledgments}

We thank Marc Genton, Tilmann Gneiting, Alex Lenkoski, Roman Schefzik and Bert Van Schaeybroeck for sharing their thoughts and expertise.  The work of Thordis L. Thorarinsdottir was supported by Statistics for Innovation, {\em sfi}$^2$, in Oslo. The work of Michael Scheuerer was supported by the German Federal Ministry of Education and Research, in the framework of the extramural research program of Deutscher Wetterdienst.

\bibliography{calibration}

\section*{Appendix}

We consider here the special case where the components of the forecast curves are independent while the components of the observation curves are fully dependent (i.e.\ identical).  As usual, we also assume that all curves are independent.  Let $X_{ik}$ be the random variable corresponding to the $k$th component of curve $i$, $f$ its density and $F$ its cumulative distribution function for $k = 1, \ldots, d$ and $i = 1, \ldots, m$.The ranks $\textup{rank}(X_{mk})$ are then also random quantities and can be written as
\[
\textup{rank}(X_{mk}) = \sum_{i=1}^m \mathbbm{1}\{X_{ik} \leq X_{mk}\}.
\]
Under the above assumptions, these quantities are uniformly distributed on $\{1,\ldots,m\}$, and hence have mean $\frac{m+1}{2}$ and variance $\frac{m^2-1}{12}$ for every $k\in\{1,\ldots,d\}$.  The relations in \eqref{eq:expected prerank} then easily follow. 

To establish the expressions for $\Var(\rho_S^{\textup{bd}}({\mathbf X}_i))$ and $\Var(\rho_S^{\textup{a}}({\mathbf X}_i))$ for the pre-rank functions in \eqref{eq:bd prerank simple} and \eqref{eq:average prerank}, respectively, we proceed as follows. For $i = 1, \ldots, m-1$, we assume that 
\[
\Var\big(\rho_S^{\textup{bd}}({\mathbf X}_i)\big) \approx \frac{1}{d^2} \sum_{k=1}^d \Var \big( (m+1) \textup{rank} (X_{ik}) - \textup{rank}(X_{ik})^2 \big),  
\]
and similar for $\Var(\rho_S^{\textup{a}}({\mathbf X}_i))$. An application of Faulhaber's formula, 
\[
\sum_{i=1}^m i^3 = \frac{m^2 (m+1)^2}{4}, \quad \sum_{i=1}^m i^4 = \frac{m (m+1) (2m+1) (3m^2 + 3m - 1)}{30},
\]
then leads to the results in \eqref{eq:var avg forecast} and \eqref{eq:var bd forecast}.

Since $X_{mk}$ takes the same value (almost surely) for all $k$, we can write $X_{mk}=X_{m\ast}$. By using the independence assumptions (between curves on the one hand and components of the forecast vectors on the other hand) we obtain for $k\neq k'$ 
\begin{eqnarray*}
 \E\big(\textup{rank}(X_{mk}) \textup{rank}(X_{mk'})\big) 
  & = & \sum_{i=1}^m\sum_{i^\prime=1}^m P\big(X_{ik}\leq X_{mk}, X_{i^\prime k'}\leq X_{mk'}\big) \\
  & = & 1 + \frac{2(m-1)}{2} + \sum_{i=1}^{m-1}\sum_{i^\prime=2}^{m-1} P\big(X_{ik}\leq X_{m\ast}, X_{i^\prime k'} \leq X_{m\ast}\big) \\
  & = & m + \frac{(m-1)^2}{3}.
\end{eqnarray*}
The last equality uses the independence of $X_{ik}, X_{i^\prime k'}$, and $X_{m\ast}$ which permits the calculation of the joint probability via Fubini,
\[
 P\big(X_{ik}\leq X_{m\ast}, X_{i^\prime k'} \leq X_{m\ast}\big) 
 = \int_{-\infty}^\infty \big(F(y)\big)^2 f(y) dy = \int_0^1 y^2 dy = \frac{1}{3}.
\]
This finally yields
\[
 \Cov\big(\textup{rank}(X_{mk}), \textup{rank}(X_{mk'})\big) = m + \frac{(m-1)^2}{3} - \frac{(m+1)^2}{4} = \frac{(m-1)^2}{12}, \qquad k\neq k',
\]
from which we obtain equation \eqref{eq:var avg obs}. 

The results for the band depth ranking in \eqref{eq:var bd obs} addtionally require the calculation of 
\begin{align*}
 \E\big(\textup{rank}(X_{mk}) \textup{rank}(X_{mk'})^2 \big) & =  \frac{3m^3+4m^2+3m+2}{12}, \\
 \E\big(\textup{rank}(X_{mk})^2 \textup{rank}(X_{mk'})^2 \big) & =  \frac{6m^4+9m^3+8m^2+3m+4}{30} 
\end{align*}
which are obtained in a similar manner (but with many more cases).

\end{document}